\documentclass[epj]{svjour}
\usepackage{amscd}
\usepackage{amssymb}
\usepackage{amsmath}
\usepackage[T1]{fontenc}
\usepackage[latin1]{inputenc}
\usepackage{graphicx}
\usepackage{graphics}
\usepackage{color}
\usepackage{latexsym}
\usepackage{amsfonts}
\usepackage{openone}
\def\be{\begin{equation}}
\def\ee{\end{equation}}
\def\ba{\begin{eqnarray}}
\def\ea{\end{eqnarray}}
\def\>{\rangle}
\def\<{\langle}
\def\n{\nonumber}

\def\sc{\scriptsize}

\begin{document}
\title{Optimization of cooling load in quantum self-contained fridge based on endoreversible approach}
\author{Ilki Kim\inst{1}\thanks{\emph{e-mail}: hannibal.ikim@gmail.com} \and Soumya S. Patnaik\inst{2}}
\institute{Center for Energy Research and Technology, North Carolina
A$\&$T State University, Greensboro, NC 27411, U.S.A. \and Aerospace
Systems Directorate, Air Force Research Laboratory, Dayton, OH
45433, U.S.A.}
\date{\today}
%
\abstract{We consider a quantum self-contained fridge consisting of
three qubits interacting with three separate heat reservoirs,
respectively, and functioning without any external controls.
Applying the methods of endoreversible thermodynamics, we derive
explicit expressions of cooling load versus efficiency of this
fridge, which demonstrate behaviors of trade-off between those two
quantities and thus enable to discuss the thermoeconomic
optimization of performance. We also discuss a possibility for the
amplification of cooling load briefly in a simple modification from
the original architecture of fridge.
}
\authorrunning{I. Kim and S. S. Patnaik}
\titlerunning{Optimization of quantum self-contained fridge $\cdots$}
\maketitle
%
\section{Introduction}\label{sec:introduction}
The issue of cooling has become a considerably important subject in
response to the arguable prediction that continued greenhouse gas
emissions at the current rate would give rise to further warming and
then many significant changes in the global climate system in future
\cite{IPC07}. Numerous cooling mechanisms and techniques have been
developed so far in different industrial systems of refrigeration
\cite{HAN12}. As an underlying formalism of those mechanisms, a
fridge driven by three heat reservoirs, without any extra work
sources, has been extensively studied \cite{AND83,SIE90}. In fact,
this architecture of fridge has attracted widespread industrial
interest due to its own potential that various forms of low-quality
energy such as the waste heat produced in industrial and biological
processes could be practically used as driving sources of
refrigeration (``energy harvesting'').

The methods of finite-time thermodynamics have been applied to
non-equilibrium thermodynamic processes observed in fridge
formalisms \cite{AND83,SIE90}. They have been able to determine the
performance bounds and optimal paths of those processes, primarily
in terms of cooling load corresponding to output power of
work-producing engines, in the context of optimized energy
management \cite{SIE94,CWU96,BER99,SAH99,KOD03,QIN05}. Also, as a
model of describing non-equilibrium processes, endoreversible
thermodynamics has been typically applied to a fridge in such a way
that its working substance is operated reversibly (``infinitely''
slowly) by an external driving in the Carnot limit while it
interacts with heat reservoirs by exchanging heat {\em
irreversibly}. Therefore, this model is highly useful for
performance study of the three-reservoir fridges operating at (real)
finite rates.

A big challenge in thermodynamics has arisen with the
miniaturization of devices achieved from the remarkable advancement
of technology, particularly in the low-temperature regime, where
quantum effects are dominant (``quantum thermodynamics'') and the
cooling mechanism is a significant issue for high-level performance
of quantum devices \cite{MAH04}. While most of other proposals for
nanoscale quantum fridges have been made to be driven by external
controls (being, however, highly nontrivial to implement
experimentally with required precision), on the other hand the
so-called ``self-contained'' fridge is operating, remarkably, merely
by incoherent interactions with three heat reservoirs at different
temperatures \cite{LIN10,POP10,SKR11,BRU12,CHE12,MAT10}, thus being
similar to the aforementioned classical model of three-bath fridges
in terms of the underlying formalism. Also, it has been found that
this quantum model is universal in its efficiency (``coefficient of
performance''), depending, i.e., upon the temperatures of baths
only, but not upon any other details such as the coupling strengths
between system and bath, notably at both the well-known Carnot limit
and being away from it \cite{BRU12}.

Motivated by this formal similarity, it is interesting to compare
both external-work-free fridges (classical and quantal) and apply
the endoreversible approach also to this quantum model in order to
look into the finite-time thermodynamic aspects therein, which have
not yet been discussed extensively. In this paper we intend to
investigate, specifically, the cooling load versus coefficient of
performance (COP), and its optimization of the quantum
self-contained fridge, as well as some relevant issues, which would
provide a foundational guidance for performance enhancement in
different types of nanoscale fridges, and thus insights into the
fundamental mechanism of overcoming various drawbacks observed in
the cooling methods at macroscopic level.

The general layout of this paper is the following. In Sect.
\ref{sec:basics} we briefly review the characteristics of quantum
self-contained fridge, needed for our discussion. In Sect.
\ref{sec:design} we derive an explicit expression of cooling load
versus COP, which enables to discuss the thermoeconomic optimization
of performance. In Sect. \ref{sec:optimization} we discuss the
cooling load versus a design parameter characterizing the fridge
architecture. We also take into consideration a simple modification
of fridge architecture in order to explore a possibility of the
amplification of cooling load. Finally we give the concluding
remarks of this paper in Sect. \ref{sec:conclusions}.

\section{Characteristics of self-contained fridge}\label{sec:basics}
The system under consideration consists of three qubits, whose
energy spacings are given by $E_a$, $E_b$, and $E_c = E_b - E_a$,
respectively, on the assumption that $E_a < E_b$ (cf.
Fig.~\ref{fig:fig1}). We are interested in building a cooling
machine by merely contacting the three qubits one-on-one to three
separate heat baths at different temperatures, initially prepared as
($T_1, T_2, T_3$) with $T_1 > T_2
> T_3$; thereby the ``hot'' bath
${\mathcal B}_1$ at $T_1$ provides heat $Q_1$ into this system which
can then extracts heat $Q_3$ steadily from the ``cold'' bath
${\mathcal B}_3$ at $T_3$. From \cite{BRU12}, it is true that such a
system can function as a fridge only when the ``biggest'' qubit with
$E_b$ is in contact with the ``intermediate'' bath ${\mathcal B}_2$
at $T_2$. Therefore, from now on, let the qubits denoted by ($E_1,
E_2, E_3$), with $E_1 < E_2$ and $E_3 = E_2 - E_1$, be in contact
with baths (${\mathcal B}_1, {\mathcal B}_2, {\mathcal B}_3$),
respectively. Moreover, let each qubit $E_j$ be initially in
equilibrium with ${\mathcal B}_j$, before the actual cooling process
occurs. And we introduce the ratio of energy spacing, $\alpha =
E_1/E_2 < 1$ as a design parameter characterizing the fridge
architecture.

By construction that no extra work is put into the system for
driving, its total energy remains unchanged during the cooling
process, and it is only possible to observe energy exchanges between
the same energy levels inside that system. To make it function as a
fridge indeed, it is simply required that before both qubits $E_1$
and $E_2$ interact with qubit $E_3$ to be cooled, the condition
$P(0,1,0) < P(1,0,1)$ be met \cite{LIN10}. Here, $P(0,1,0)$ denotes
the probability of $E_1$ being in its ground state, $E_2$ being in
its excited state, and $E_3$ being in its ground state, as well as
the probability $P(1,0,1)$ follows similarly. This is equivalent to
$e^{-E_2/T_2} < e^{-E_1/T_1}\cdot e^{-E_3/T_3}$ (with
$k_{\mbox{\tiny B}} = 1$), which will easily reduce to the
inequality $T_3 > T_v > 0$, expressed in terms of the virtual
temperature \cite{BRU12}
\begin{equation}\label{eq:virtual_temp1}
    T_v\, =\, \frac{1 - \alpha}{1/T_2 - \alpha/T_1}\,,
\end{equation}
being independent of $T_3$. Therefore, as long as this inequality
condition is met, the fridge functions. Fig. \ref{fig:fig2}
demonstrates the behaviors of $T_v$ versus $\alpha$.

It has been shown \cite{POP10,SKR11} that each of heat exchanged
between qubit and bath satisfies in the steady-state the ratio
\begin{equation}\label{eq:ratio-of-heat}
    Q_1 : Q_2 : Q_3\; =\; E_1 : E_2 : E_3\,.
\end{equation}
The COP of this quantum fridge then equals \cite{BRU12}
\begin{equation}\label{eq:COP1}
    \eta_{\mbox{\scriptsize fr}}^{(q)}\; :=\; \frac{Q_3}{Q_1}\; =\; \frac{E_3}{E_1}\; =\; \frac{T_2^{-1} - T_1^{-1}}{T_v^{-1} -
    T_2^{-1}}\,.
\end{equation}
In comparison, the well-known Carnot value $\eta_{\mbox{\tiny C}}$
is given by (\ref{eq:COP1}) with $T_v \to T_3$, which is valid for
the classical three-bath fridges as well; note that the value
$\eta_{\mbox{\tiny C}}$ of fridges may be greater than unity.
Similarly, by replacing $T_v$ by $T_3$ in (\ref{eq:virtual_temp1})
and solving for $\alpha$, we can obtain the Carnot value
\begin{equation}\label{eq:carnot_value1}
    \alpha_{\mbox{\tiny C}}\; =\; \frac{T_3^{-1} - T_2^{-1}}{T_3^{-1} - T_1^{-1}} < 1\,.
\end{equation}
Therefore, only the region of $\alpha_{\mbox{\tiny C}} < \alpha < 1$
is allowed for cooling process, as shown in Fig. \ref{fig:fig2}. At
the minimum $\alpha_{\mbox{\tiny C}}$, the fridge in fact stops
functioning (in the frame of finite-time thermodynamics) due to the
fact that $T_3 \ngtr T_v$. It is noted in (\ref{eq:carnot_value1})
that with $T_3$ becoming lower, the working region of $\alpha$
shrinks. Also, if $T_2 \to T_3$, then $\eta_{\mbox{\tiny C}} \to
\infty$ and $\alpha_{\mbox{\tiny C}} \to 0^+$, equivalent to $E_1
\to 0$, which implies the breakdown of this cooling system.

We should pay special attention to the case of $\alpha = 1/2$, in
which an additional channel of heat transport is open, i.e., a
direct heat flow from $E_1$ to $E_3$ due to the fact that $T_1 >
T_3$, being not available at all when $\alpha \ne 1/2$. This is in
fact a heat flow in the opposite direction to the cooling process.
Therefore, in this case, we need an extra check-up for bringing the
overall cooling process true: Since heat flux from one side to the
other one is proportional to temperature difference between both
sides (cf. details in Sect. \ref{sec:design}), it is required that
the temperature difference $T_3 - T_v$ for cooling be greater than
$T_1 - T_3$ (on the assumption that heat conductances in both
directions are the same). This condition easily reduces to the
inequality, $0 > T_1^2 - 2\,T_3\,T_1 + T_2\,T_3 := f(T_1)$, being
quadratic in $T_1$. However, the discriminant of $f(T_1)$, given by
$D = 4\,T_3\,(T_3 - T_2) < 0$, indicates that $0 \ngtr f(T_1)$
indeed for all $T_1$. As a result, this system cannot function as a
fridge at $\alpha = 1/2$, in addition to $\alpha = 0, 1$.

We close this section by reminding that when the actual cooling
proceeds, each qubit $E_j$ is not in equilibrium with ${\mathcal
B}_j$ any longer, thus being not at $T_j$.

\section{Cooling load versus fridge efficiency}\label{sec:design}
We first consider a classical model of fridge driven by three heat
baths. By applying the first law of thermodynamics to its cyclic
process in the (non-equilibrium) steady-state, we easily obtain
\begin{equation}\label{eq:endo1}
    Q_1 - Q_2 + Q_3\, =\, 0\,.
\end{equation}
All of irreversibility during the cooling process may be split into
two parts \cite{KOD03}; one is the external irreversibility
occurring in (system-bath) heat exchange, resulting from the
temperature differences between baths and working substance of the
fridge, while the other is the internal irreversibility resulting
from all entropy-producing dissipations inside the working
substance, say, friction, mass transfer, etc. By applying the second
law to the working substance, we obtain
\begin{equation}\label{eq:endo1-1}
    \frac{Q_1}{T_{i1}} - \frac{Q_2}{T_{i2}} + \frac{Q_3}{T_{i3}}\, \geq\, 0\,,
\end{equation}
where the symbol $T_{ij}$ with $j = 1, 2, 3$ denotes internal
effective temperature of the working substance's subsystem being in
direct interaction with bath ${\mathcal B}_j$. Here, the condition
of external irreversibility requires that $T_{i1} < T_1$, and
$T_{i2}
> T_2$, as well as $T_{i3} < T_3$. To simplify our discussion, we apply endoreversible
thermodynamics to (\ref{eq:endo1-1}) by neglecting the internal
irreversibility, thus the inequality (\ref{eq:endo1-1}) reducing to
the equality
\begin{align}\tag{\ref{eq:endo1-1}a}\label{eq:endo1-2}
    \frac{Q_1}{T_{i1}} - \frac{Q_2}{T_{i2}} + \frac{Q_3}{T_{i3}}\, =\, 0\,,
\end{align}
which will be in consideration from now on. Combining
(\ref{eq:endo1}) and (\ref{eq:endo1-2}) then allows us to have the
equality of fridge efficiency (COP)
\begin{equation}\label{eq:endo2}
    \eta_{\mbox{\scriptsize fr}}\, =\, \frac{Q_3}{Q_1}\, =\, \frac{T_{i2}^{-1} - T_{i1}^{-1}}{T_{i3}^{-1} -
    T_{i2}^{-1}}\,.
\end{equation}
If heat exchange is carried out infinitely slowly, then the
steady-state reduces to the quasi-static state with $T_j \leftarrow
T_{ij}$ and so the Carnot value $\eta_{\mbox{\tiny C}} \leftarrow
\eta_{\mbox{\scriptsize fr}}$. We will below apply those results of
endoreversible thermodynamics to the quantum self-contained fridge,
say, by identifying $\eta_{\mbox{\scriptsize fr}}^{(q)}$ with
(\ref{eq:endo2}). In fact, this does not any harm since this quantum
fridge already has no source of internal irreversibility at all, due
to its architecture.

As a quantity of finite-time thermodynamics, heat flux is now
considered during the cooling process, given by $\dot{Q} = k\,(T_h -
T_l)$ obeying the Newtonian linear law, in which $k$ denotes a heat
conductance, as well as $T_h$ and $T_l$ are high and low
temperatures, respectively \cite{SIE90}. From this, it follows that
$\dot{Q}_1 = k_1\,(T_1 - T_{i1})$, and $\dot{Q}_2 = k_2\,(T_{i2} -
T_2)$, as well as $\dot{Q}_3 = k_3\,(T_3 - T_{i3})$. We now
introduce the specific cooling load for the quantum fridge as
\begin{equation}\label{eq:endo3}
    {\mathcal L}_3 := \dot{Q}_3/K\,,
\end{equation}
where the total heat conductance $K = k_1 + k_2 + k_3$. Then it is
straightforward to transform (\ref{eq:endo3}) into
\begin{equation}\label{eq:endo4}
    {\mathcal L}_3 =
    \left(\frac{\dot{Q}_1}{\dot{Q}_3}\,\frac{1}{T_1 - T_{i1}} +
    \frac{\dot{Q}_2}{\dot{Q}_3}\,\frac{1}{T_{i2} - T_2} + \frac{1}{T_3 -
    T_{i3}}\right)^{-1}
\end{equation}
which, with the aid of (\ref{eq:endo1}) and (\ref{eq:endo2}),
reduces to
\begin{equation}\label{eq:endo5}
    {\mathcal L}_3(\eta_{\mbox{\scriptsize fr}}) = \left\{\frac{1}{\eta_{\mbox{\scriptsize fr}}\,(T_1 - T_{i1})} +
    \frac{(1/\eta_{\mbox{\scriptsize fr}}) + 1}{T_{i2} - T_2} + \frac{1}{T_3 -
    T_{i3}}\right\}^{-1}\,,
\end{equation}
then interpreted as a function of a given COP
$\eta_{\mbox{\scriptsize fr}}$.

Now we are interested in optimizing this expression of specific
cooling load in order to see its steady-state behaviors. Therefore
we consider
\begin{equation}\label{eq:endo5_1}
    \tilde{\mathcal L}_3\, =\, {\mathcal L}_3 + \lambda\,\left(\eta_{\mbox{\scriptsize fr}} -
    \frac{y - x}{1 - y}\right)\,,
\end{equation}
where $\lambda$ is a Lagrangian multiplier, as well as $x :=
T_{i3}/T_{i1}$ and $y := T_{i3}/T_{i2}$. By requiring that $\partial
\tilde{\mathcal L}_3/\partial x = 0$, and $\partial \tilde{\mathcal
L}_3/\partial y = 0$, as well as $\partial \tilde{\mathcal
L}_3/\partial T_{i3} = 0$, followed by algebraic manipulation, we
can determine the optimized values of three effective temperatures
as
\begin{subequations}
\begin{eqnarray}
    T_{i1} &=& \frac{1}{2}\,\left\{T_1\, +\, \frac{T_2\,(1 + \eta_{\mbox{\scriptsize fr}}\cdot T_1/T_3)}{1 + \eta_{\mbox{\scriptsize fr}}}\right\}\label{eq:endo6_1}\\
    T_{i2} &=& \frac{1}{2}\,\left\{T_2\, +\, \frac{T_3\,(1 + \eta_{\mbox{\scriptsize fr}})}{T_3/T_1 + \eta_{\mbox{\scriptsize fr}}}\right\}\\
    T_{i3} &=& \frac{1}{2}\,\left\{T_3\, +\, \frac{T_2\,(T_3/T_1 + \eta_{\mbox{\scriptsize fr}})}{1 + \eta_{\mbox{\scriptsize fr}}}\right\}\,,\label{eq:endo6_3}
\end{eqnarray}
\end{subequations}
expressed in terms of three given bath temperatures and a given COP
only. Substituting (\ref{eq:endo6_1})-(\ref{eq:endo6_3}) into
(\ref{eq:endo5}), we can finally obtain
\begin{equation}\label{eq:endo7}
    {\mathcal L}_3(\eta_{\mbox{\scriptsize fr}})\, =\,
    \frac{T_2}{4\,(1 + \eta_{\mbox{\scriptsize fr}})}\, -\, \frac{T_3}{4\,(1 + \eta_{\mbox{\scriptsize fr}}\cdot T_1/T_3)}\,
    -\, \frac{(T_2 - T_3)}{4}\,.
\end{equation}
It is easy to confirm that ${\mathcal L}_3(\eta_{\mbox{\tiny C}}) =
0$ and ${\mathcal L}_3(0) = 0$ indeed, as required. Further,
requiring $\partial {\mathcal L}_3/\partial \eta_{\mbox{\scriptsize
fr}} = 0$ will give the maximum value of (\ref{eq:endo7}),
\begin{equation}\label{eq:endo8}
    {\mathcal L}_m\; =\; \frac{(\sqrt{T_1\,T_2} - T_3)^2}{4\,(T_1 - T_3)} - \frac{(T_2 -
    T_3)}{4}\,,
\end{equation}
evaluated at $\eta_m = (1 - \sqrt{T_2/T_1})/(\sqrt{T_1\,T_2}/T_3 -
1)$. Fig. \ref{fig:fig3} demonstrates behaviors of the specific
cooling load versus COP for various ($T_1,T_2,T_3$); here we may
regard ${\mathcal L}_3/{\mathcal L}_m$ as a measure of
irreversibility.

Next, with the help of (\ref{eq:endo6_1})-(\ref{eq:endo6_3}), the
heat fluxes can be determined as
\begin{subequations}
\begin{eqnarray}
    \dot{Q}_1\, =\, \frac{k_1\,A}{2\,(1 + \eta_{\mbox{\scriptsize fr}})} &,&
    \dot{Q}_2\, =\, \frac{k_2\,(T_3/T_1)\,A}{2\,(T_3/T_1 + \eta_{\mbox{\scriptsize fr}})}\\
    \dot{Q}_3 &=& \frac{k_3\,(T_3/T_1)\,A}{2\,(1 + \eta_{\mbox{\scriptsize
    fr}})}\,,\label{eq:cooling_flux1}
\end{eqnarray}
\end{subequations}
where $A := T_1 - T_2 - \eta_{\mbox{\scriptsize fr}}\,(T_2 -
T_3)\,T_1/T_3$. Now we require that the heat conductances satisfy
\begin{equation}\label{eq:heat_conductances1}
    k_2/k_1 = 1 + \eta_{\mbox{\scriptsize fr}}\,T_1/T_3\;\; ,\;\; k_3/k_1 = \eta_{\mbox{\scriptsize
    fr}}\,T_1/T_3\,.
\end{equation}
This will immediately yield
\begin{equation}\label{eq:heat_conductances2}
    \dot{Q}_1 = \dot{Q}_2/(1 + \eta_{\mbox{\scriptsize fr}})\;\;,
    \;\; \dot{Q}_3 = \dot{Q}_2\,\eta_{\mbox{\scriptsize fr}}/(1 + \eta_{\mbox{\scriptsize
    fr}})\,,
\end{equation}
and $\dot{Q}_2 = k_2\,(A/2)\,(1 + \eta_{\mbox{\scriptsize
fr}}\,T_1/T_3)^{-1}$ with $k_2 \propto E_2$. With the help of
(\ref{eq:COP1}), it is then easy to show that Eq.
(\ref{eq:heat_conductances2}) consists with (\ref{eq:ratio-of-heat})
indeed. It is stressed again that heat conductances ($k_1,k_2,k_3$)
in the steady state can be uniquely determined (up to constant) for
a given COP and the initial conditions, ($T_1,T_2,T_3$) and
($E_1,E_2$). Fig. \ref{fig:fig4} shows behaviors of the heat fluxes
$(\dot{Q}_1,\dot{Q}_2,\dot{Q}_3)$ versus COP.

In comparison, we consider the remaining specific heat loads,
${\mathcal L}_1 = \dot{Q}_1/K$ and ${\mathcal L}_2 = \dot{Q}_2/K$.
Along the same lines as applied above for ${\mathcal
L}_3(\eta_{\mbox{\scriptsize fr}})$, it is straightforward to obtain
${\mathcal L}_1(\eta_{\mbox{\scriptsize fr}}) = {\mathcal
L}_3/\eta_{\mbox{\scriptsize fr}}$ and ${\mathcal
L}_2(\eta_{\mbox{\scriptsize fr}}) = {\mathcal L}_3\cdot(1 +
\eta_{\mbox{\scriptsize fr}})/\eta_{\mbox{\scriptsize fr}}$, with
the same optimized values of effective temperatures as given in
(\ref{eq:endo6_1})-(\ref{eq:endo6_3}). Then, both of heat loads will
yield exactly the same behaviors of
$(\dot{Q}_1,\dot{Q}_2,\dot{Q}_3)$ as those derived from ${\mathcal
L}_3(\eta_{\mbox{\scriptsize fr}})$, shown in Fig. \ref{fig:fig4}.
This fact is no surprise since the optimization process in
(\ref{eq:endo5_1}) was taken into consideration in order to find
uniquely the steady-state behaviors of heat fluxes.

It is also instructive to apply to this quantum fridge the
thermoeconomic criterion introduced in \cite{SAH99}, then given by
\begin{equation}\label{eq:thermoeconomics1}
    {\mathcal F}_c\; :=\; \frac{\dot{Q}_3}{C_t}\; =\; (a\,\eta_{\mbox{\scriptsize fr}}^{-1} +
    b\,{\mathcal L}_3^{-1})^{-1}\,.
\end{equation}
Here the total cost $C_t$ consists of both energy consumption cost
$C_e = a\,\dot{Q}_1$ and investment cost $C_i = b\,K(E_1,E_2,E_3)$
being assumed to be linearly proportional to the system size $E_j$
each, with proportionality coefficients $a$ and $b$; e.g., in case
of the waste heat in industry to be used as the input energy for
fridges, then $C_e \ll C_i$. The case of $a=0$ and $b=1$ gives
${\mathcal F}_c \to {\mathcal L}_3$ while the opposite case of $a=1$
and $b=0$, on the other hand, gives ${\mathcal F}_c \to
\eta_{\mbox{\scriptsize fr}}$. This behavior shows a trade-off
between ${\mathcal L}_3$ and $\eta_{\mbox{\scriptsize fr}}$ in the
context of thermoeconomics. By substituting (\ref{eq:endo7}) into
(\ref{eq:thermoeconomics1}), we can observe those behaviors of
${\mathcal F}_c(\eta_{\mbox{\scriptsize fr}})$ for various values of
$a$ and $b$, as demonstrated in Fig. \ref{fig:fig5}.

Moreover, we note that the total COP in (\ref{eq:endo2}) may be
rewritten as $\eta_{\mbox{\scriptsize
he}}^{(1,2)}\,\eta_{\mbox{\scriptsize fr}}^{(2,3)}$ (cf.
\cite{SKR11}), in which both efficiencies are explicitly given by
$\eta_{\mbox{\scriptsize he}}^{(1,2)} = (T_{i1} - T_{i2})/T_{i1}$ of
an endoreversible heat engine ($E_1,E_2$) and
$\eta_{\mbox{\scriptsize fr}}^{(2,3)} = T_{i3}/(T_{i2} - T_{i3})$ of
an endoreversible Carnot-type fridge ($E_2,E_3$). In comparison, it
is worthwhile to focus on the sub-fridge which is exactly driven by
the output work $W_{1,2}\,(= W_{2,3})$ of the engine ($E_1,E_2$) and
then produces by itself the same cooling flux as $\dot{Q}_3$ in
(\ref{eq:heat_conductances2}). Along the same lines as applied for
(\ref{eq:endo3})-(\ref{eq:endo7}), its cooling flux in the steady
state can be found as \cite{CWU96}
\begin{equation}\label{eq:endo-Carnot-fridge1}
    \dot{Q}_3\left(\eta_{\mbox{\scriptsize fr}}^{(2,3)}\right)\; =\;
    \frac{k_{2,3}\,\left\{T_3 - T_2\cdot\eta_{\mbox{\scriptsize fr}}^{(2,3)} \left(1 +
    \eta_{\mbox{\scriptsize fr}}^{(2,3)}\right)^{-1}\right\}}{(\sqrt{k_{2,3}/k_3} +
    1)^2}\,,
\end{equation}
expressed in terms of ($T_2,T_3$) and $\eta_{\mbox{\scriptsize
fr}}^{(2,3)}$; here the symbol $k_{2,3}$ denotes the heat
conductance (to be determined) of (partial) heat flux directly from
this sub-fridge to the sink ${\mathcal B}_2$ such that
$\dot{Q}_{2,3} = k_{2,3}\,(T_{i2} - T_2) = \dot{W}_{2,3} +
\dot{Q}_3$. By equating (\ref{eq:endo-Carnot-fridge1}) to
(\ref{eq:heat_conductances2}), it is straightforward to evaluate
$k_{2,3}(\eta_{\mbox{\scriptsize
he}}^{(1,2)},\eta_{\mbox{\scriptsize fr}}^{(2,3)})$ explicitly. From
this, we can also determine the heat conductance $k_{1,2} = k_2 -
k_{2,3}$ of heat flux directly from the engine ($E_1,E_2$) to the
sink ${\mathcal B}_2$ such that $\dot{Q}_{1,2} = k_{1,2}\,(T_{i2} -
T_2) = \dot{Q}_1 - \dot{W}_{1,2}$. Then it is obvious to confirm
that $\dot{W}_{1,2} = \dot{W}_{2,3}$.

\section{Optimal design of the fridge}\label{sec:optimization}
Now we are interested in physically designing the quantum fridge
functioning optimally. To do so, it is needed to select the
optimized value of design factor $\alpha$. With the help of
(\ref{eq:ratio-of-heat}) and (\ref{eq:endo1}), we can easily get
$\eta_{\mbox{\scriptsize fr}} = (1/\alpha) - 1$. Substituting this
into (\ref{eq:endo7}) will allow us to have the specific cooling
load versus $\alpha$, given by
\begin{eqnarray}\label{eq:endo11}
    {\mathcal L}_3(\alpha) &=& \frac{T_2}{4}\,(\alpha - 1)\,
    +\, \frac{1}{4\,(T_3^{-1} - T_1^{-1})}\, +\n\\
    && \frac{(T_1/T_3 - 1)^{-2}/4}{\alpha/T_1 - 1/(T_1 - T_3)}\,.
\end{eqnarray}
Fig. \ref{fig:fig6} shows this, as well as $\eta_{\mbox{\scriptsize
fr}}(\alpha)$ in comparison; the maximum of ${\mathcal L}_3(\alpha)$
is located at $\alpha_m = (1 - T_3/\sqrt{T_1\,T_2})/(1 - T_3/T_1)$
while ${\mathcal L}_3(\alpha_{\mbox{\tiny C}}) = 0$ and ${\mathcal
L}_3(1) = 0$. The first subregion of $\alpha_m \leq \alpha < 1$
exactly corresponds to $0 < \eta \leq \eta_m$ in Fig. 3 while the
second subregion of $\alpha_{\mbox{\tiny C}} < \alpha \leq \alpha_m$
to $\eta_m \leq \eta < \eta_{\mbox{\tiny C}}$ in which a trade-off
between ${\mathcal L}_3$ and $\eta_{\mbox{\scriptsize fr}}$ is
explicitly seen. The behaviors of heat fluxes
$(\dot{Q}_1,\dot{Q}_2,\dot{Q}_3)$ versus $\alpha$ are also plotted
in Fig. \ref{fig:fig7}.

Two comments deserve here. First, we easily find that ${\mathcal
L}_3(\alpha) \ne {\mathcal L}_3(1-\alpha)$. In fact, as pointed out
in Sect. \ref{sec:basics}, we should pay extra attention to the
symmetric point $\alpha = 1/2$, where ${\mathcal L}_3(1/2) =
\{2\,T_1\,T_3/(T_1 + T_3) - T_2\}/8$. We first consider the case
that $\alpha_{\mbox{\tiny C}}
> 1/2$. Then it follows from (\ref{eq:carnot_value1}) that ${\mathcal L}_3(1/2) < 0$,
thus showing that $\alpha = 1/2$ already lies out of the working
region. When $\alpha_{\mbox{\tiny C}} < 1/2$, on the other hand, it
follows that ${\mathcal L}_3(1/2) > 0$, but we should exclude
$\alpha = 1/2$ in designing this quantum fridge.

Secondly, in \cite{BRU12} the analysis was conducted for given $E_1$
and $E_2$, thus $\alpha$ being merely a constant, which then proved
that the fridge functions only for the case that the temperature
($T_3$) of an external system to be cooled is lower than the
resultant machine temperature $T_v > 0$ expressed in terms of
$\alpha$ (as well as $T_1$ and $T_2$). On the other hand, in our
analysis the focus has been mainly taken on building a fridge to be
driven by initially given temperatures ($T_1, T_2, T_3$), by
determining its working region in terms of control parameter
$\alpha$, and then optimizing its performance.

As a next step of performance study, it is also interesting to make
a simple modification from the original architecture
$(E_1,E_2,E_3)$, in order to explore a possibility for the
amplification of cooling flux. For a given total system size $E_t =
E_1 + E_2 +E_3$, let two identical sub-fridges be available, given
by $(E_1/2,E_2/2,E_3/2)_a$ and $(E_1/2,E_2/2,E_3/2)_b$, each being
in contact with baths $({\mathcal B}_1,{\mathcal B}_2,{\mathcal
B}_3)$, respectively. We now consider two different cases. In the
first case, let both sub-fridges be not allowed to interact with
each other. Then, only the heat input filtered by a single energy
spacing $(E_1/2)$ from the bath ${\mathcal B}_1$ can be used for
extraction of heat from ${\mathcal B}_3$. It is easy to show that
the COP of each subsystem is $(\eta_{\mbox{\scriptsize fr}})_a =
(\eta_{\mbox{\scriptsize fr}})_b$, being identical to
(\ref{eq:endo2}). It is also true that each of cooling flux is
$(\dot{Q}_3)_a = (\dot{Q}_3)_b$, and so the flux of the total system
is $2\,(\dot{Q}_3)_a$, being identical to (\ref{eq:cooling_flux1}),
too. Therefore, we do not observe any amplification of cooling flux
from the total system in this case of modification.

On the other hand, in the second case, let the two sub-fridges
 be allowed to interact, whose combined system is denoted by
$\{(E_1/2,E_1/2), (E_2/2,E_2/2), (E_3/2,E_3/2)\}$. First, this total
system contains $2 \times 2 \times 2$ identical cooling channels
(``type 1''). It is then easy to show that the 8 fridges are
identical in COP, in fact given in (\ref{eq:endo2}). And the cooling
flux of each sub-fridge is $(\dot{Q}_3)_s \propto (1/4)\,(E_3/2)$,
thus the flux obtained from the 8 channels altogether exactly
amounting to (\ref{eq:cooling_flux1}). However, in this case, there
is an extra cooling channel (``type 2'') which comes out from the
heat input filtered by energy spacing $(E_1)$ from ${\mathcal B}_1$.
Then, the {\em virtual} qubits $(E_1,E_2,E_3)$ of this channel are
initially prepared at the virtual temperatures $(T_1,T_2,T_3)$,
respectively. This channel is obviously not available in the first
case at all. From the above two types of channels altogether, the
total cooling flux of this second case is accordingly amplified as
$2\,\dot{Q}_3$ (factor 2) while the total COP still equals
(\ref{eq:endo2}). Then, at the optimized value $\alpha_m$, the
maximally amplified cooling flux is obtained from (\ref{eq:endo11}).
Here it was assumed that first, the fridge-bath coupling strengths
are equal for both types, and thus in dealing with COP and cooling
flux of the total system, the two types of channels equally
contribute to averaging those two quantities; secondly, there is no
additional cooling channel which satisfies matching the energy
spacing such as $E_1 + E_2 = E_3$. This result of amplification in
cooling flux can be generalized for a more complicated modification
of architecture as long as all sub-fridges are allowed to interact,
and so multi-channels of cooling are available.

\section{Concluding remarks}\label{sec:conclusions}
In summary, we studied the performance of quantum self-contained
fridges by applying endoreversible thermodynamics. We analyzed
behaviors of cooling load versus coefficient of performance, and
then their optimization, in terms of the design parameter. This
verified that a trade-off between those two quantities exists
indeed, also in this external-work-free quantum fridge. In doing so,
we uniquely determined heat conductances in the steady state for
given initial conditions, which enabled our result to consist with
the previous findings in references. We also studied a possibility
for the amplification of cooling load briefly in a simple
modification from the original architecture of fridge. As a next
step, it is suggested to take into consideration the heat exchange
between fridge and baths obeying the non-Newtonian law, as well as a
more complicated modification of architecture in which the
multi-channels of cooling are available.

As a result, our approach will contribute to providing a
foundational guidance for the thermoeconomic optimization of
performance for nano-scale fridges functioning in the quantum
thermodynamic regime. This also suggests that engineering methods
can apply to the study of fundamental science, which has not
extensively been carried out thus far.

\section*{Acknowledgments}
One of the authors (I. K.) thanks G. Mahler (Stuttgart), S. Deffner
(LANL), G. J. Iafrate (NC State) and J. Kim (KIAS) for helpful
remarks. He appreciates the support from the AF Summer Faculty
Fellowship Program during his visit to AFRL, Wright-Patterson AF
Base, Ohio. He acknowledges financial support provided by the U.S.
Army Research Office (Grant No. W911NF-15-1-0145).

\onecolumn{
\begin{figure}[htb]
\thicklines \unitlength=1mm
\begin{picture}(137,150)(0,50)
\put(15,110){\dashbox{0.7}(15,15){\null}}
\put(22.5,110){\vector(0,1){15}} \put(22.5,125){\vector(0,-1){15}}
\put(15,109){\rule{15\unitlength}{1\unitlength}}
\put(21.5,105){${\displaystyle T_1}$}
\put(17,116.6){\shortstack[l]{${\displaystyle E_1}$}}
\put(40,110){\dashbox{0.7}(15,25){\null}}
\put(40,109){\rule{15\unitlength}{1\unitlength}}
\put(42,121){\shortstack[l]{${\displaystyle E_2}$}}
\put(47.5,110){\vector(0,1){25}} \put(47.5,135){\vector(0,-1){25}}
\put(46.5,105){${\displaystyle T_2}$}
\put(65,110){\dashbox{0.7}(15,10){\null}}
\put(65,109){\rule{15\unitlength}{1\unitlength}}
\put(67,114.4){\shortstack[l]{${\displaystyle E_3}$}}
\put(72.5,110){\vector(0,1){10}} \put(72.5,120){\vector(0,-1){10}}
\put(71.5,105){${\displaystyle T_3}$} \put(15,110){\line(1,0){65}}
\put(15,125){\line(1,0){15}} \put(40,135){\line(1,0){15}}
\put(65,120){\line(1,0){15}}
\put(15,125){\rule{15\unitlength}{1\unitlength}}
\put(40,135){\rule{15\unitlength}{1\unitlength}}
\put(65,120){\rule{15\unitlength}{1\unitlength}}
\end{picture}
\centering\hspace*{-4.1cm}{\caption{\label{fig:fig1}}}
\end{figure}
Fig.~\ref{fig:fig1}: A schematic description of the fridge
consisting of three qubits whose energy spacings are given by
$(E_a,E_b,E_c) = (E_1,E_2,E_3)$ with $E_3 = E_2 - E_1$. Those qubits
$(E_1,E_2,E_3)$ are in contact with three separate heat baths
$({\mathcal B}_1,{\mathcal B}_2,{\mathcal B}_3)$ at temperatures
$(T_1,T_2,T_3)$, respectively.
\newpage
\begin{figure}[htb]
\centering\hspace*{-2.5cm}\vspace*{4cm}{
\includegraphics[scale=0.75]{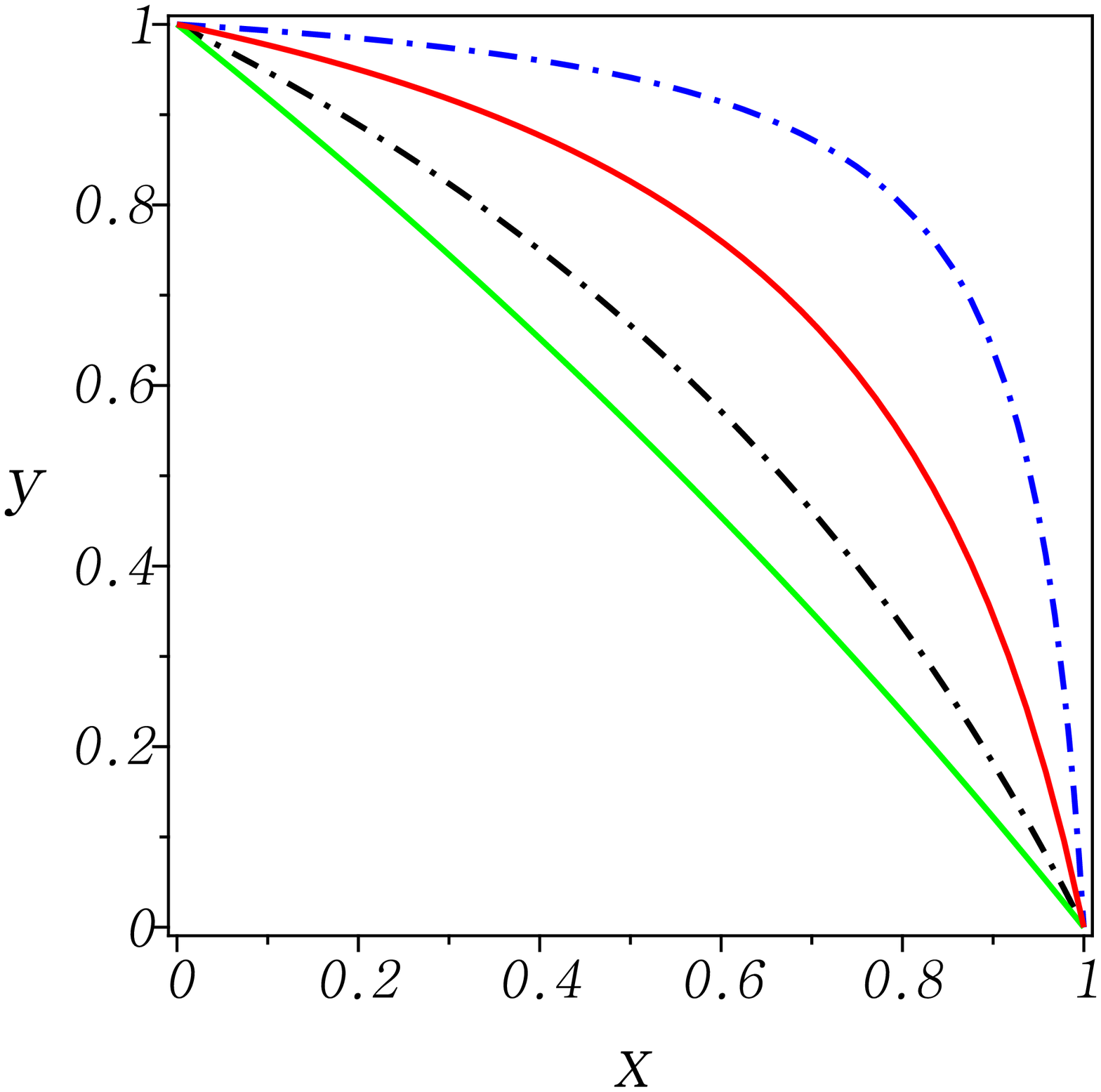}
\caption{\label{fig:fig2}}}
\end{figure}
Fig.~\ref{fig:fig2}: (Color online) Dimensionless virtual
temperature $y = T_v/T_2$ in (\ref{eq:virtual_temp1}) versus
dimensionless quantity $x = \alpha$. From the bottom,
``low-temperature'' regime: (I) (solid, green): $T_1 = 5$ and $T_2 =
1$, (II) (dashdot, black): $T_1 = 2$ and $T_2 = 1$;
``high-temperature'' regime: (III) (solid, red): $T_1 = 19$ and $T_2
= 15$, (IV) (dashdot, blue): $T_1 = 16$ and $T_2 = 15$.
\newpage
\begin{figure}[htb]
\centering\hspace*{-2.5cm}\vspace*{4cm}{
\includegraphics[scale=0.75]{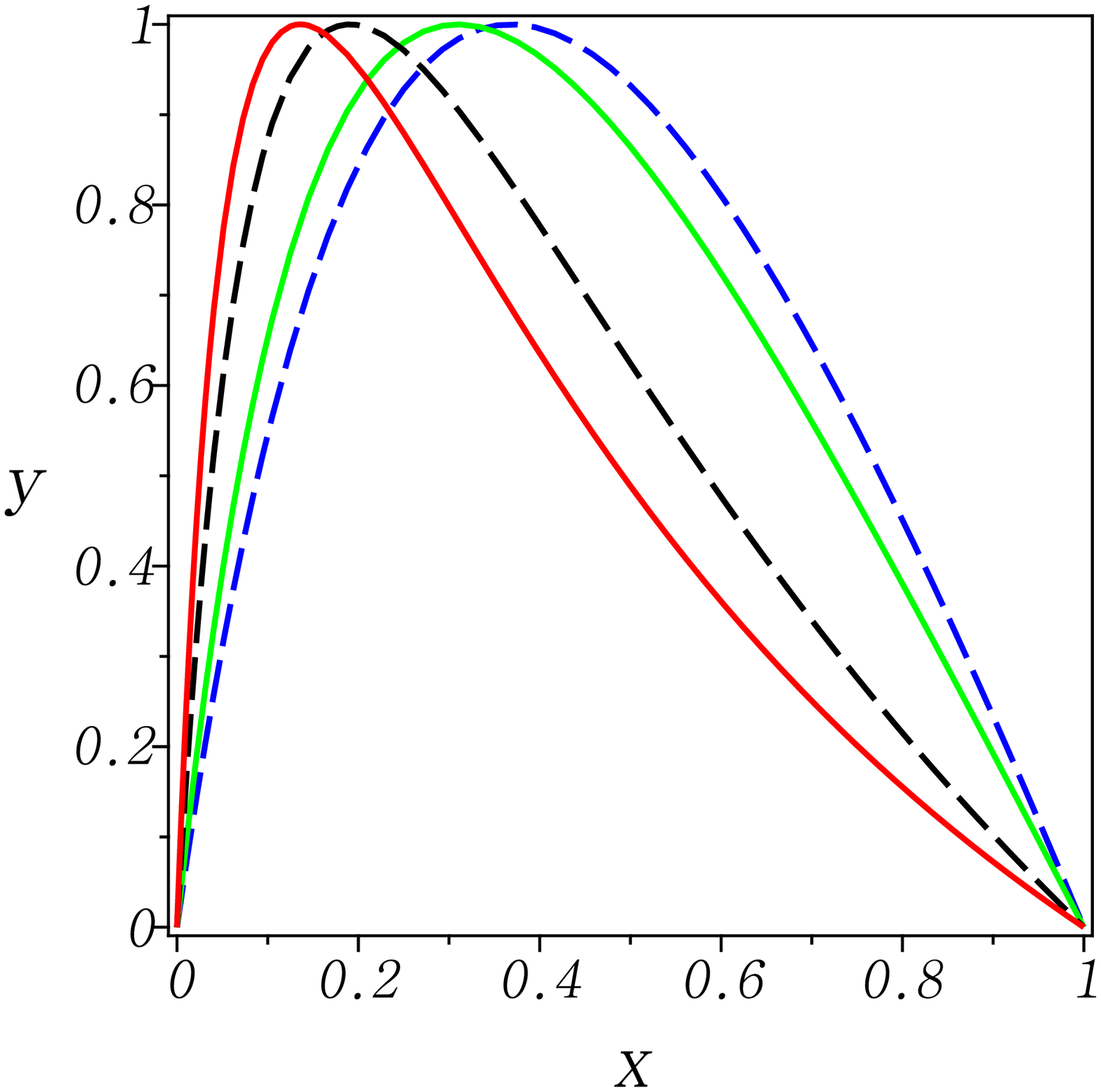}
\caption{\label{fig:fig3}}}
\end{figure}
Fig.~\ref{fig:fig3}: (Color online) Normalized specific cooling load
$y = {\mathcal L}_3/{\mathcal L}_m$ (as a ``measure of
irreversibility'') given in (\ref{eq:endo7}) versus normalized
efficiency $x = \eta_{\mbox{\scriptsize fr}}/\eta_{\mbox{\tiny C}}$.
Let $t_1 := T_1/T_2$ and $t_3 := T_3/T_2$. From the left in maximum
value position $x_m$, (I) (solid, red): $t_1 = 3$ and $t_3 = 0.8$;
(II) (dash, black): $t_1 = 2$ and $t_3 = 0.8$; (III) (solid, green):
$t_1 = 3$ and $t_3 = 0.3$; (IV) (dash, blue): $t_1 = 2$ and $t_3 =
0.3$.
\newpage
\begin{figure}[htb]
\centering\hspace*{-2.5cm}\vspace*{4cm}{
\includegraphics[scale=0.75]{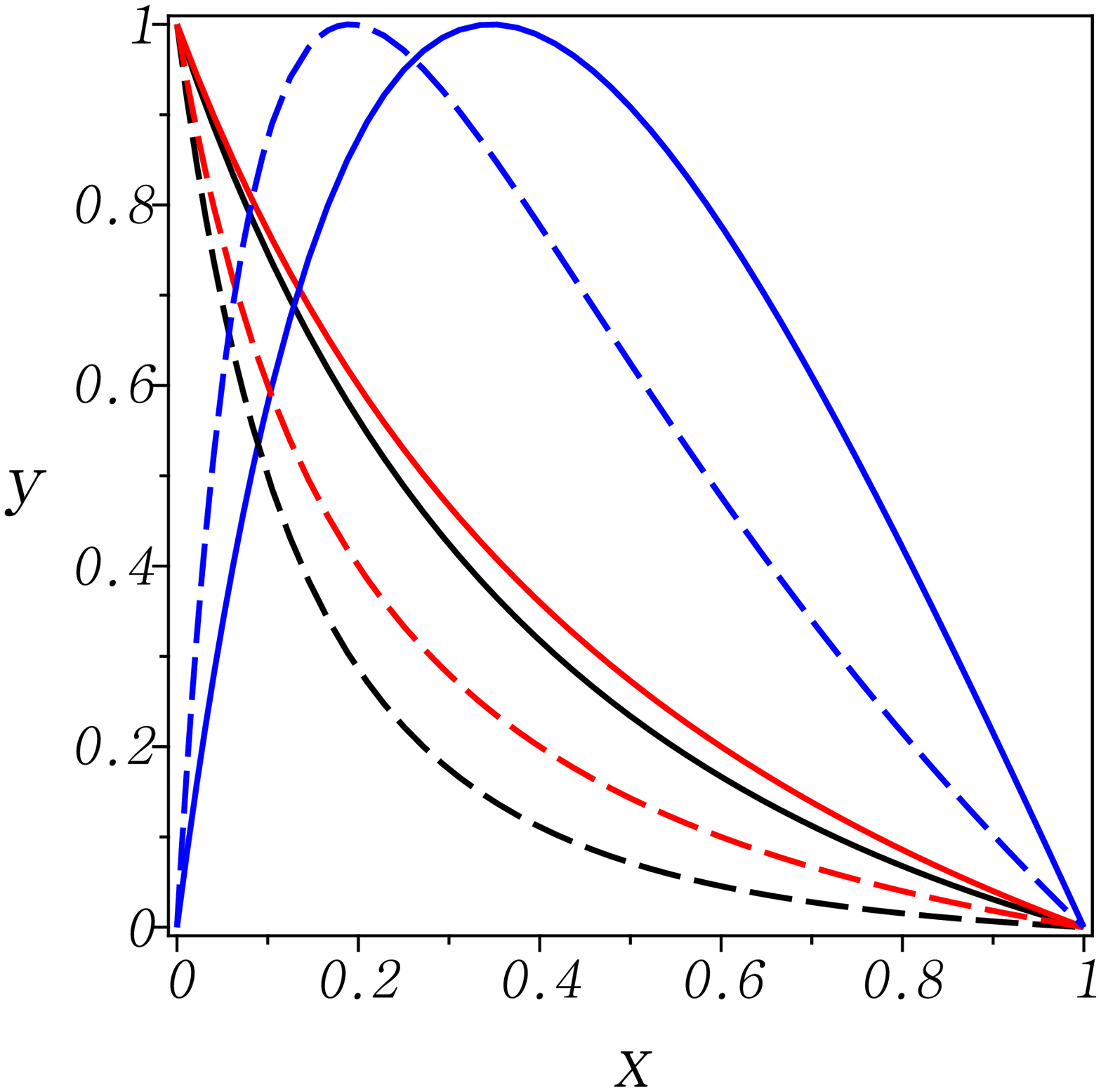}
\caption{\label{fig:fig4}}}
\end{figure}
Fig.~\ref{fig:fig4}: (Color online) Normalized heat flux $y =
\dot{q}_j := \dot{Q}_j/(\dot{Q}_j)_m$ in
(\ref{eq:heat_conductances2}) versus normalized efficiency $x =
\eta_{\mbox{\scriptsize fr}}/\eta_{\mbox{\tiny C}}$, where $j =
1,2,3$, and $(\dot{Q}_j)_m$ denotes maximum value of $\dot{Q}_j$.
Let $t_1 := T_1/T_2$ and $t_3 := T_3/T_2$. Dashed curves represent
the case of $(t_1, t_3) = (2, 0.8)$: From the bottom at $x = 1/2$,
(I) $\dot{q}_1$ (black); (II) $\dot{q}_2$ (red); (III) $\dot{q}_3$
(blue) being the cooling flux. Solid curves represent the case of
$(t_1, t_3) = (2, 0.4)$ and follow along the same lines.
\newpage
\begin{figure}[htb]
\centering\hspace*{-2.5cm}\vspace*{4cm}{
\includegraphics[scale=0.75]{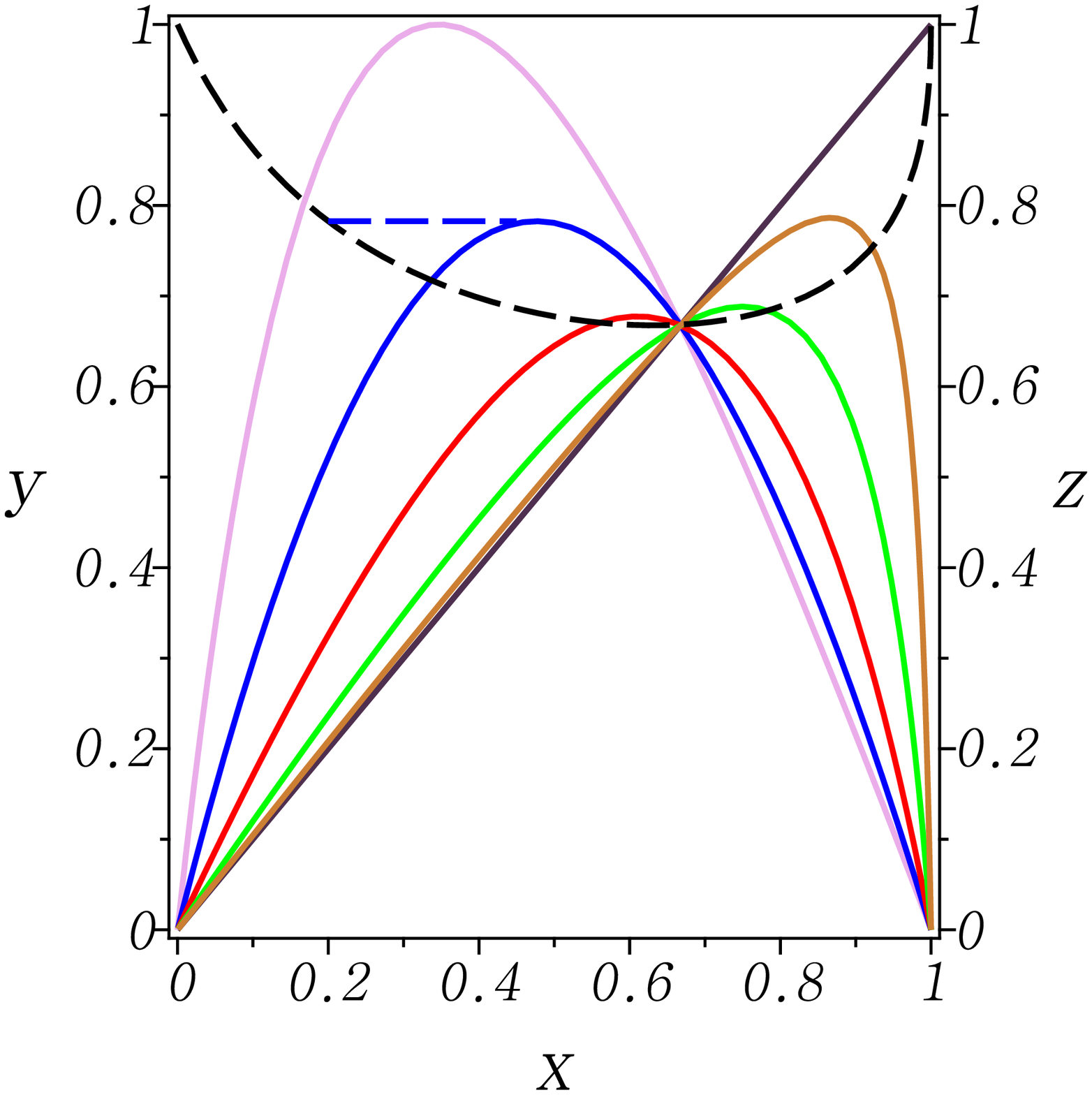}
\caption{\label{fig:fig5}}}
\end{figure}
Fig.~\ref{fig:fig5}: (Color online) Solid lines: Dimensionless
thermoeconomic criterion $y = {\mathcal F}_c = \{a'\,x^{-1} +
b'\,({\mathcal L}_3/{\mathcal L}_m)^{-1}\}^{-1}$ in
(\ref{eq:thermoeconomics1}) versus normalized efficiency $x =
\eta_{\mbox{\scriptsize fr}}/\eta_{\mbox{\tiny C}}$, where $a' =
a/\eta_{\mbox{\tiny C}}$ and $b' = b/{\mathcal L}_m$. We set $b' = 1
- a'$, and $(t_1, t_3) = (2, 0.4)$. From the bottom at $x = 0.9$:
(I) $a' = 0$ (plum) representing ${\mathcal L}_3/{\mathcal L}_m$;
(II) $a' = 0.2$ (blue); (III) $a' = 0.5$ (red); (IV) $a' = 0.8$
(green); (V) $a' = 0.95$ (gold); (VI) $a' = 1$ (violet) being a
straight line $y = x$. In comparison, a dashed (black) curve $z =
{\mathcal F}_m(a')$ versus $x = a'$ is also put, which shows the
maximum value of $y$ for a given $a'$; e.g., in case of $a' = 0.2$,
the curve $y$ (blue) has its maximum value of $0.78262 = {\mathcal
F}_m(0.2)$, as indicated by the segment of dash line (blue).
\newpage
\begin{figure}[htb]
\centering\hspace*{-2.5cm}\vspace*{4cm}{
\includegraphics[scale=0.75]{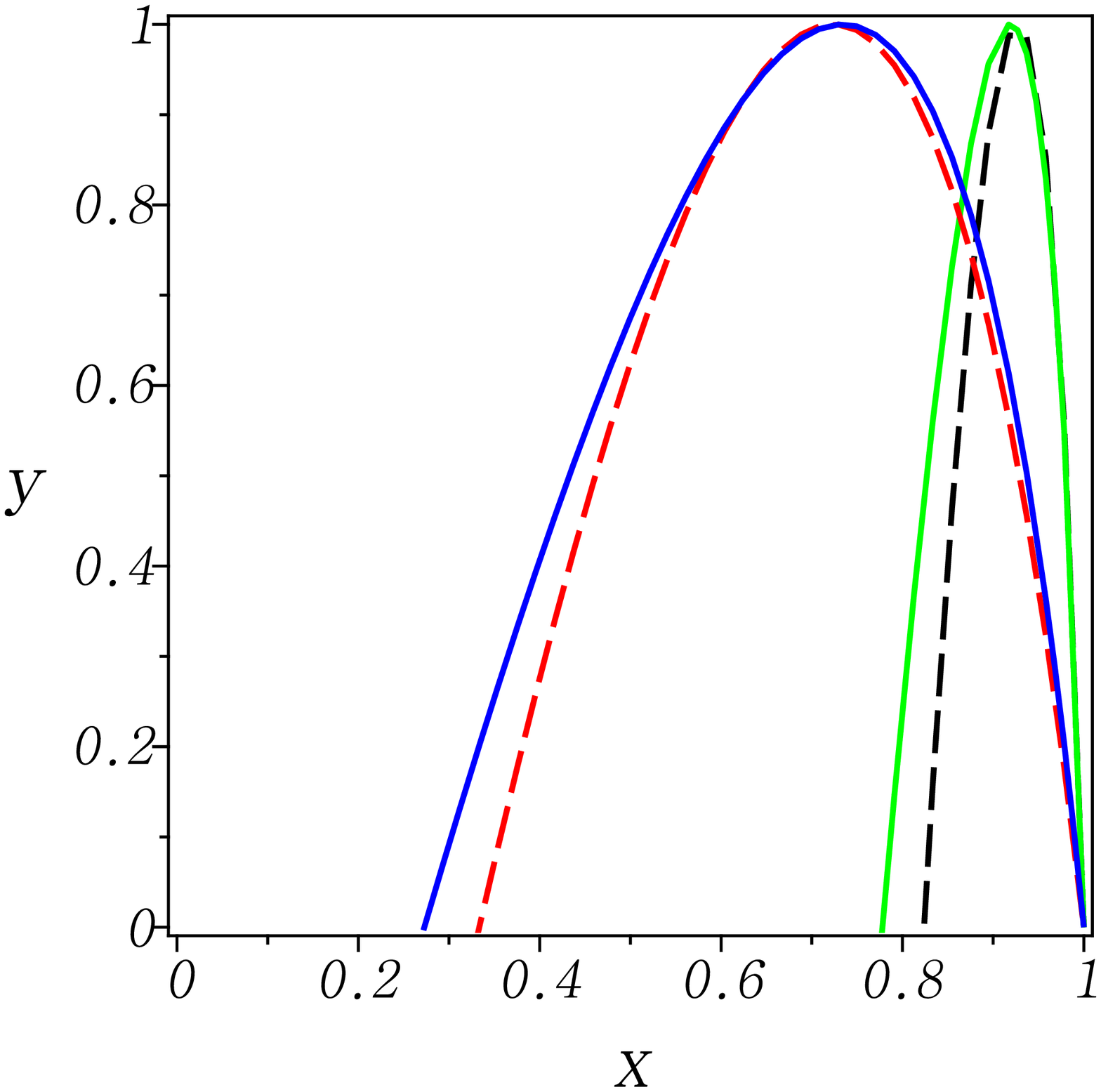}
\caption{\label{fig:fig6}}}
\end{figure}
Fig.~\ref{fig:fig6}: (Color online) Normalized specific cooling load
$y = {\mathcal L}_3/{\mathcal L}_m$ in (\ref{eq:endo11}) versus $x =
\alpha$. From the right in position of $x = \alpha_{\mbox{\tiny
C}}\,(\ne 1)$ at which $y = 0$: (I) (dash, black): $t_1 = 2$ and
$t_3 = 0.3$, (II) (solid, green): $t_1 = 3$ and $t_3 = 0.3$, (III)
(dash, red): $t_1 = 2$ and $t_3 = 0.8$, (IV) (solid, blue): $t_1 =
3$ and $t_3 = 0.8$.
\newpage
\begin{figure}[htb]
\centering\hspace*{-2.5cm}\vspace*{4cm}{
\includegraphics[scale=0.75]{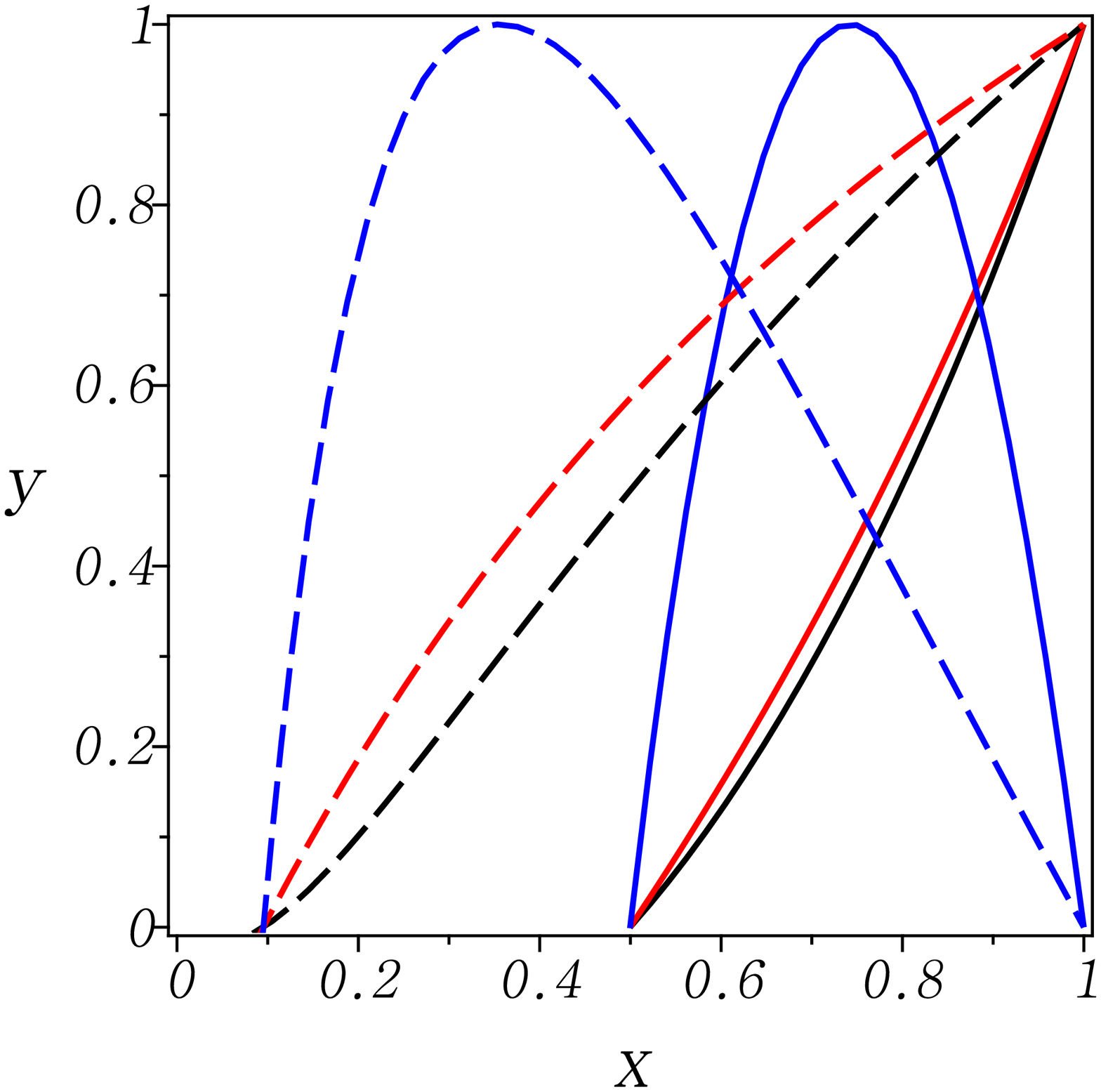}
\caption{\label{fig:fig7}}}
\end{figure}
Fig.~\ref{fig:fig7}: (Color online) Normalized heat flux $y =
\dot{q}_j := \dot{Q}_j/(\dot{Q}_j)_m$ (to be derived from
(\ref{eq:endo11})) versus $x = \alpha$, where $j = 1,2,3$, and
$(\dot{Q}_j)_m$ denotes maximum value of $\dot{Q}_j$. Solid curves
represent the case of $(t_1, t_3) = (2, 2/3)$ equivalent to
$\alpha_{\mbox{\tiny C}} = 1/2$: From the bottom at $x = 0.6$, (I)
$\dot{q}_1$ (black); (II) $\dot{q}_2$ (red); (III) $\dot{q}_3$
(blue). Dashed curves represent the case of $(t_1, t_3) = (2, 0.95)$
and follow similarly.
}
\end{document}